\documentclass[prd,aps]{revtex4}
\usepackage{graphicx}

\begin{document}
\title{A physical distinction between a covariant and a non-covariant
reduction process in relativistic quantum theories.}
\author{Rodolfo Gambini$^{1}$ and Rafael A. Porto$^{1}$}
\affiliation {1. Instituto de F\'{\i}sica, Facultad de Ciencias,
Igu\'a 4225, esq. Mataojo, Montevideo, Uruguay.}
\date{March 27th 2002}

\begin{abstract}
Causality imposes strong restrictions on the type of operators
that may be observables in relativistic quantum theories. In fact,
causal violations arise when computing conditional probabilities
for certain partial causally connected measurements using the
standard non covariant procedure. Here we introduce another way of
computing conditional probabilities, based on an intrinsic
covariant relational order of the events, which differs from the
standard one when this type of measurements are included. This
alternative procedure is compatible with a wider and very natural
class of operators without breaking causality. If some of these
measurements could be implemented in practice, as predicted by our
formalism, the non covariant, conventional approach should be
abandoned. Furthermore, the description we promote here would
imply a new physical effect where interference terms are
suppressed as a consequence of the covariant order in the
measurement process.
\end{abstract}

\maketitle

As it has been shown by many
authors\cite{A-A,Sorkin,Preskill,Preskill2}, causality imposes
strong restrictions on the type of operators that may be
observables in a measurement process. These restrictions arise, as
we are about to see, when one considers certain particular
arrangements composed by partial causally connected regions where
conditional measurements take place. This type of causal
connection appears when the regions of space-time where the
quantum states are subject to a measurement process are partially
time-like and partially space-like separated. While some operators
are admissible in the relativistic domain, many others are not
allowed by the standard formalism. This conclusion may be derived
from two basic hypotheses, the {\it minimally disturbing
hypothesis} \cite{Sorkin} and the conventional Bloch notion for
ordering the events in the relativistic domain. The previously
mentioned hypothesis assumes that the conditional probability
calculus for a given set of observables related to some operators
of the theory, can be obtained, without introducing the
experimental devices within the theory, with a wide range of
accuracy. In other words, that it is enough to consider the system
decoupled from the experimental devices, with a given probability
formula and a reduction postulate for the quantum state after each
observable is measured on certain space-like region. On the other
hand, Bloch's approach consists in choosing an arbitrary
Lorentzian reference system and hence:{\it "...the right way to
predict results obtained at $C$ is to use the time order that the
three regions $A,B,C$ have in the Lorentz frame that one happens
to be using"}\cite{Bloch}. Although here we are going to use the
{\it minimally disturbing hypothesis}, we are going to use a
different, covariant notion of order\cite{nos1}, and its
corresponding reduction postulate, which we shall see, implies
different predictions in the case of non-local partial causally
connected measurements, though coincides otherwise. We are going
to show in what follows some particular cases where this physical
distinction is manifest.\\

As it has been recently advocated the {\it minimally disturbing
hypothesis} we are going to adopt here could be avoided in the
non-relativistic domain if there is decoherence in a particular
basis of the device's degrees of freedom, once taking the trace on
the environment\cite{zurek}. There, the {\it usual} probability
distributions are recovered and the above mentioned hypothesis, is
therefore based on more physical grounds. This has attracted a lot
of attention, now, where the border between the classical and the
quantum world is under current observation.\\ If one looks for an
extension of this program to the relativistic domain one faces new
problems. First of all, there is not a well defined relativistic
quantum theory of single particles and one must extend quantum
mechanics to field theories. On the other hand there is not a
unique time order for the conditional probability calculus to be
used and causal problems require further study.\\ Within these
considerations, the most natural approach has been to hold the
{\it minimally disturbing hypothesis} and use the microcausality
property of quantum field operators, that is the commutation of
operators associated to space-like separated regions, to show that
causality is preserved for any local measurement and therefore,
that the non covariant order is harmless. Concerning non-local
measurements one could study to what extent this approach is
physically viable and hope that in a not too far future we will be
able to understand better the relativistic extension of the
measurement process. Along this line it has been shown as we said
above that certain non-local operators would not be measurable
quantities. In other words, one wouldn't be able to measure them
keeping track just of the state of the quantum field system within
the traditional non covariant approach we mentioned before.
Nevertheless a final theory is still lacking and the observable
character of those operators is yet controversial. To our
knowledge there is not a single approach to overcome the
measurement problem in quantum field theory where the causal and
space-time properties of the measurement process are matched
together. The analogous decoherence effect has not been fully
understood and it has not been shown that the non covariant order
for conditional probabilities is harmless when including the
device within the theory though it is natural to think that the
microcausality property will play an essential role.\\ It is then
meaningful to explore the possibility of holding the {\it
minimally disturbing hypothesis} as a first attempt to understand
the causal versus the non-local aspects of the measurement process,
by extending our relational covariant approach in order to include
a wider class of admissible non-local operators. We will discuss
the physical consequences of this extension later on.\\

In a previous set of papers \cite{nos1,nos2,nos3} we have
introduced a covariant realistic description of quantum states and
the reduction process in relativistic quantum mechanics and
relativistic quantum field theories, something which has been
studied by many authors in different contexts
\cite{A-A,Bloch,Esp,Tommas,Suarez,Griff}. We have shown that it is
possible to extend a realistic description to the relativistic
domain where a quantum state may be considered as a relational
object that characterizes the disposition of the system for
producing certain events with certain probabilities among a given
intrinsic set of alternatives. To understand this notion of
reality better let us remember what Omnes used to ask about
physical processes: {\em ``tell me a story"} \cite{Omnes}. The
actors of this story are the building blocks of physics, for
instance electrons, quarks etc. But there is a problem with these
actors, there is not even a  {\it play} until they act. This is
what quantum mechanics has taught us, there are not properties
before measurements. So, the image of the electron crossing for
instance a Geiger counter is just that, a picture. One can think,
nevertheless, that there is some kind of reality in this {\it
play}, a relational one\cite{nos1,nos2,nos3}. The actors exist
because they have the capacity of {\it talking} to the {\it
audience}. Properties are the result of the interaction. In that
sense, {\em a system is given by the set of its behaviors with
respect to others} . An isolated system does not have properties
or attributes, since all the ``properties" result from its
interaction with other systems. It is important to remark that
this is a strongly objective description in the sense of
D'Espagnat and it does not make any reference to operations
carried out by human observers \cite{Esp}. In order to extend this
point of view into the relativistic domain one begins introducing
an intrinsic order for the set of alternatives on the measurement
process as follows: let us denote by $A_{R,u}$ the instrument
associated to a spacelike region $R$ whose four-velocity is $u$.
We started by introducing the following partial order: the
instrument $A_{{R_1},{u_1}}$ precedes $A_{{R_2},{u_2}}$ if the
region $R_2$ is totally contained in the forward light cone of
$R_1$. Let us suppose that ${A^0}_{{R_0},{u_0}}$ precedes all the
others. In other words, we assume that all the detectors are
inside the forward light cone coming from this initial condition.
That would be the case, for instance, for the instrument that
prepares the initial state $s=0$ in the EPR(B) experiment. In this
way one can introduce a strict order without any reference to a
Lorentz time. Define $S^1$ as the set of instruments that are
preceded only by $A^0$. Define $S^2$ as the set of instruments
that are preceded only by the set $S^1$ and $A^0$. In general,
define $S^i$ as the set of instruments that are preceded by the
sets $S^j$ with $j<i$ and $A^0$. Notice that any couple of
elements in $S^i$ is separated by space-like intervals. This
procedure defines a covariant order based on the causal structure
induced by the devices involved in the measurement
process\footnote{As one can immediately notices, the initial
condition has a deep relevance in the construction of the
covariant alternatives. In many cases the preparation of the
system is central for the determination of the initial condition.
As one already knows, a quantum system involve entangled objects,
therefore in a complete quantum theory one has to take the whole
universe as the system. There, the relational point of view is the
only way for describing the evolution. In this domain, a quantum
object may not have a natural beginning beyond the big bang. If
one is describing a particular portion of the universe within a
given time interval, then one can consider a partial initial
condition given by a particular set of events that contain the
forthcoming alternatives in the forward light cone. Hence, one
naturally falls into a sort of statistical mixture, as it is the
case in non-complete measurements.}. The crucial observation is
that all the alternatives on $S^i$ can be considered as
``simultaneous" for the decision process of the quantum state.\\
Contrary to the non covariant approach, we shall see that an
extension of this description to the case of partial causal
connections allows us to include a wider class of causal
operators. It is then very important to understand whether or not
this intrinsic order is physically relevant.\\

Let us introduce the experimental arrangement shown in figure 1.
Let us suppose, following Sorkin\cite{Sorkin}, that the
measurement set-up is composed by two regions, $A$,$C$, and one
intermixed partial causally connected measurement on region $B$,
associated to values of certain Heisenberg observables. We shall
denote ${P^A}_a$,${P^B}_b$, ${P^C}_c$ their corresponding
Heisenberg projectors, where the upper labels represent the region
and the lower ones the eigenvalues of the corresponding operators.

\begin{figure}[ht]
\begin{center}
\includegraphics*[width=6cm]{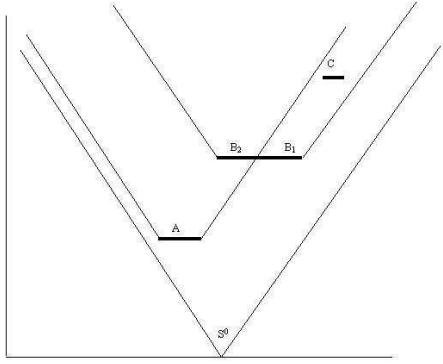}
\caption{Sorkin's arrangement with an intermixed partial causally
connected measurement.}
\end{center}
\end{figure}

Let us suppose that the measurement in region $B$ admits the
decomposition of the associated operator in two partial operators
related with $B_1$ and $B_2$. That is, the field operator
associated to the region $B$ can be put as a function of two new
operators related to the portions of $B$ such that $B=B_1\bigcup
B_2$. Let us denote the respective eigenvalues $b_1$ and $b_2$,
and suppose that the total result $b$ on $B$ is extensive in the
sense that $b=f(b_1,b_2)$. For instance, let us call $(O^1,O^2)$
the operators associated to the partial regions $(B_1,B_2)$ and
$O$ the operator associated to $B$. Therefore, $f(O^1,O^2)=O$ is
the functional relation between them. Notice that this hypothesis
includes a wide class of operators since we do not make any extra
assumption on this functional relation. We shall describe these
partial projectors by $P^{B_1}_{b_1},P^{B_2}_{b_2}$. Then, due to
microcausality\footnote{Here we use that
$[P^{B_1}_{b_1},P^{B_2}_{b_2}]=0$ which implies that we can
diagonalize both self-adjoint operators, $O^1,O^2$, in the same
orthonormal basis. The vectors in this basis are thus also
eigenvectors for the operator $O=f(O^1,O^2)$ with eigenvalues
$b=f(b_1,b_2)$, where $b_i$ are the eigenvalues for the $O_i$
operators. The following property is a natural decomposition for
the ${P^B}_b$ projector.}:

\begin{equation}
\sum_{b_1}\sum_{b_2}\delta(b-f(b_1,b_2)){P^{B_2}}_{b_2}{P^{B_1}}_{b_1}=
{P^B}_b . \label{proj}
\end{equation}

It is important to stress that the decomposition of the projector
associated to the measurement on region $B$ is not part of the
experimental arrangement that includes independent measurements on
the portions $B_1$ and $B_2$, but a consequence of the relational
intrinsic order as we are going to show below.\\ Let us analyze
this experiment using Bloch's notion of order to define the
sequence of options $S^1$,$S^2$, $S^3$, and the corresponding
reduction processes followed by a quantum system. It is clear that
following Bloch one would get $S^1=A<S^2=B<S^3=C$. Hence, one
immediately notices that the $A$ measurement affects the $B$
measurement and also the $B$ measurement affects the $C$
measurement. Consequently, one should expect that the $A$
measurement would affect the $C$ measurement, leading to
information traveling faster than light between $A$ and $C$, which
are space-like separated regions. One can prove this result as
follows: let us suppose that the state of the system was prepared
by a initial measurement, that precedes the whole arrangement,
whose density operator we denote by $\rho_0$. Now, the probability
of having the result $a,b,c$ in the regions $A,B,C$ given the
initial state $\rho_0$ is, using Wigner's formula within Bloch's
approach:

\begin{eqnarray}
&{\cal P^{B}}(a,b,c|\rho_0)=
Tr[{P^C}_c{P^B}_b{P^A}_a{\rho}_0{P^A}_a{P^B}_b]=&\label{probm2}\\
&\sum_{(b_1,b_2,b'_1)}\delta(b-f(b'_1,b_2))
\delta(f(b'_1,b_2)-f(b_1,b_2))
Tr[{P^{C}}_{c}{P^{B_2}}_{b_2}{P^{B_1}}_{b_1}
{P^A}_a{\rho}_0{P^A}_a{P^{B_1}}_{b'_1}{P^{B_2}}_{b_2}]& \nonumber
\end{eqnarray}

Where the sums are taken over the whole set of possible values,
and we have used the cyclic property of the trace, microcausality
and the projector character of ${P^{B_2}}_{b_2}$ \footnote{We use
that $[{P^{B_2}}_{b_2},{P^C}_c]=0$ and
${P^{B_2}}_{b_2}{P^{B_2}}_{b'_2}=\delta(b_2-b'_2){P^{B_2}}_{b_2}$}.

Now, in order to study the causal implications of this expression
we are going to suppose that non selective measurements have been
performed on $A,B$\footnote{Notice that the resulting values for
the measurement carried out on $A,B$ can not be transmitted
causally to an observer in $C$.}. Hence, the probability of having
$c$, no matter the results on $A,B$ is:

\begin{eqnarray}
&{\cal P^{B}}(\mathrm{unknown}\; a,\mathrm{unknown}\;
b,c|\rho_0)=\sum_a\sum_b{\cal
P^{B}}(a,b,c|\rho_0)=&\label{probm3}\\
&\sum_a\sum_{(b_1,b_2,b'_1)} \delta(f(b'_1,b_2)-f(b_1,b_2))
Tr[{P^{C}}_{c}{P^{B_2}}_{b_2}{P^{B_1}}_{b_1}
{P^A}_a{\rho}_0{P^A}_a{P^{B_1}}_{b'_1}{P^{B_2}}_{b_2}]& \nonumber
\end{eqnarray}

Notice that if we set $f\equiv b_1+b_2$ in equation
(\ref{probm3}), it turns out to be:

\begin{eqnarray}
&{\cal P}^{B}(\mathrm{unknown}\; a,\mathrm{unknown}\;
b,c|\rho_0)=&\label{probm4}\\ &\sum_a\sum_{(b_1,b_2,b'_1)}
\delta(b'_1-b_1) Tr[{P^{C}}_{c} {P^{B_2}}_{b_2}{P^{B_1}}_{b_1}
{P^A}_a{\rho}_0{P^A}_a{P^{B_1}}_{b'_1}]=
\sum_{b_1}Tr[{P^{C}}_{c}{P^{B_1}}_{b_1}{\rho}_0{P^{B_1}}_{b_1}]&
\nonumber
\end{eqnarray}

Where we have used that $\sum_{b_2}{P^{B_2}}_{b_2}=Id$ and $\sum_a{
{P^A}_a}=Id$. Therefore, Bloch's approach is consistent with
causality in the linear case. However, one immediately notices
that this is not a general feature. As it can be read from the
expression (\ref{probm3}) in the non linear case, this formula
breaks relativistic causality. That is due to the fact that in
general the delta function $\delta(f(b'_1,b_2)-f(b_1,b_2))$
imposes a constraint on the $b_2$ values that does not allow to
perform an independent sum by using the closure relation. Hence,
if one uses Bloch's notion for the ordering of the alternatives,
one gets faster than light signals for a wide class of operators,
those which are non linear respect to the portions of the region.
This prevents us from ignoring that the $A$ measurement has been
performed. There is no violation with respect to the $B$
observation since an observer at $C$ may be causally informed
about a measurement carried out at $B$. However, the above
analysis implies faster than light communication with respect to
the $A$ measurement since it is space-like separated from $C$.
Therefore, the requirement of causality strongly restricts the
allowed observable quantities in relativistic quantum mechanics.\\
Let us show this in detail in the particular case of a quantum
scalar field. Let us introduce the following operator for the
measurement carried out on region $B$:

\begin{equation}
{\hat
O}(t^{B})=\int_{B}\int_{B}g^{B}(x)g^{B}(y)\hat{\phi}(x,t^{B})
\hat{\phi}(y,t^{B})dxdy \label{op}
\end{equation}

where $g^{B}$ is a smooth smearing function for the field
operator, with compact support such that it is non-zero in the
region $B$. We shall assume that any projection of the state is
instantaneous at proper time $t^B$ for the local Lorentz system,
neglecting as usual the duration of the measurement
process\footnote{This idealization is possible as much as the
partial causal connection is preserved during a period of time
much longer that the duration of the measurement process itself
(decoherence time scale). That leads us to consider wide spacelike
regions for the $B$ measurements in order to retain this ideal
case. The physical effect that we are going to discuss here also
would appear for generic partial causally connected devices, but
it would need to consider space-time partial causally connected
regions instead of thin almost spacelike ones.}.

Notice now, that the above operator has a non-local
behavior with respect to the region $B$. It is easy to see that it
also implies a non linear behavior for the functional relation
$f(b_1,b_2)$. In this case we will have the partial
operators\footnote{In Ref. \cite{nos3} we have studied the
spectral decomposition for these kind of operators in the Klein
Gordon free field.}:

\begin{equation}
{\hat O_i}(t^{B})=\int_{B_i}dx g^{B}(x)\hat{\phi}(x,t^{B})\;\;
i=1,2
\end{equation}

Therefore we will get the functional relation $f=(b_1+b_2)^2$
since $O=(O^1+O^2)^2$. Now we can introduce it on equation
(\ref{probm3}) obtaining:

\begin{eqnarray}
&{\cal P^{B}}(\mathrm{unknown}\; a,\mathrm{unknown}\;
b,c|\rho_0)=&\label{probm5}\\&\sum_a\sum_{(b_1,b_2)}
\frac{1}{2(b_2+b_1)}Tr[{P^{C}}_{c}{P^{B_2}}_{b_2} {P^{B_1}}_{b_1}
{P^A}_a{\rho}_0{P^A}_a{P^{B_1}}_{b_1}]+&\nonumber\\
&+\sum_a\sum_{(b_1,b_2)}\frac{1}{2(b_2+b_1)}
Tr[{P^C}_{c}{P^{B_2}}_{b_2}{P^{B_1}}_{b_1}
{P^A}_a{\rho}_0{P^A}_a{P^{B_1}}_{-b_1-2b_2}]&\nonumber
\end{eqnarray}

where we have used that $\delta((b'_1)^2-(b_1)^2+2(b'_1-b_1)b_2)
=\frac{1}{2(b_2+b_1)}\left(\delta(b'_1-b_1) +
\delta(b'_1+b_1+2b_2)\right)$. Hence, one immediately sees that

one can not use, as in the linear case, the identity decomposition
for the $B_2$ measurement because there is not an independent sum
on $b_2$. Therefore, the standard Bloch approach doesn't allow us
to measure operators as the one defined in (\ref{op}), nor its
natural extension for an $n$-field function.\\ We shall show in
what follows that the relational approach is covariant, and
consistent with causality for the general type of operations we
are considering, while the standard expression is unacceptable as
we have seen. Let us consider again Sorkin's arrangement with the
relational intrinsic order (see figure 1). Let us start with $S^0$
and the preparation of the state in $\rho_0$. Hence we will have
$S^1=(A,B_1)$ and $S^2=(B_2,C)$. The key observation is the
following. In order to introduce the relational covariant
reduction process on this particular framework, our notion of
partial order requires to consider the measurement in region $B$
as composed by different alternatives. That is, in the case where
only a portion of the region where measurement takes place is
causally connected, one needs to decompose the region in portions
such that each part is completely inside (or outside) the forward
light cone coming from the precedent ones. Hence the
alternatives belonging to one set $S^i$ may be composed by several
parts of different instruments. In fact, a particular device could
contain parts belonging to different options. The decision process
of the quantum state for producing an observable phenomenon on
region $B$ is composed now by two new set of alternatives on $B_1$
and $B_2$. Within this new scenario it is immediately noticed that
the conditional probability formula for the above experiment
should be calculated as:

\begin{equation}
{\cal P}(a,b,c|\rho_0)={\sum}_{(b_1,b_2)} \delta(b-f(b_1,b_2))
Tr[{P^{C}}_{c}{P^{B_2}}_{b_2}{P^{B_1}}_{b_1}
{P^A}_a{\rho}_0{P^A}_a{P^{B_1}}_{b_1}{P^{B_2}}_{b_2}]\label{prob1}
\end{equation}

Where the sum is taken over the set of partial projections
compatible with the final result $b$ on region $B$. As we
mentioned before, the individuality of the whole measurement still
persists since we do not have access to any partial result
$b_1,b_2$, but only to the total result $b$ obtained on $B$ after
observation. Nevertheless, due the intrinsic relational order, the
quantum state must decide about the measurement on region $B$ in a
``non-simultaneous" set of alternatives. First the alternative
$S^1=(A,B_1)$ followed by $S^2=(B_2,C)$. This implies that the
quantum state may pass through a chain of partial decision
processes for producing the final result $b$ on $B$. The resulting
wave function collapse, associated to the registered value $b$,
should take into account this chain of partial decision processes
associated with the intrinsic covariant order. This fact is
therefore reflected in (\ref{prob1}) by the sum on the partial
projections. Notice that we are not considering an experimental
set up with independent measurements on $B_1,B_2$. In fact, it is
easy to see that both approaches would coincide for this last
case. The measurement process we are here considering only
involves the non-local measurement of $b$ on $B$ without further
information left. Hence, we should understand the process given in
(\ref{prob1}) as a consequence of the relational intrinsic order
for the case of partial causally connected non-local measurements,
rather than the result of an experiment with actual independent
measurements on the portions of region $B$. This is a very
important departure from the standard viewpoint because we are
allowing the possibility of partial decision processes, i.e.
projections, of the quantum state for producing an observable
phenomenon on region $B$ with result $b$, but without the aid of
partial local registrations. This is an inescapable consequence of
the covariant order. Therefore, (\ref{prob1}) implies a new kind
of physical process where a quantum state may be projected without
producing yet any macroscopic observable effect, but instead, as
part of a chain of decision processes which ends in the final
macroscopic result. As we are going to discuss below this effect
could be consistent with a fully Schroedinger-like description of
the measurement process.\\

It is now easy to see that this experimental
setting does not lead to causal violations for non selective
measurements on the $A,B$ regions. In order to do that, we just
perform the sum on the unknown results $a,b$ getting:

\begin{eqnarray}
&{\cal P}(\mathrm{unknown}\; a,\mathrm{unknown}\;
b,c|\rho_0)=\label{prob2}&\\ &\sum_a\sum_b{\cal P}(c,a,b|\rho_0)=
\sum_{b_1}Tr[{P^C}_c{P^{B_1}}_{b_1}\rho_0{P^{B_1}}_{b_1}]&
\nonumber
\end{eqnarray}

where we have used, making use of microcausality, the identity
decomposition for the measurement on $B_2$, and afterwards on $A$.
The final sum runs over the complete set of possible values of
$b_1$. This leads to an interesting dependence of the final
conditional probability (\ref{prob2}) on the complete set of
projections on the portion $B_1$ causally connected with $C$. This
type of {\it correlation} does not imply any incompatibility with
causality since it just informs an observer in $C$ that the $B$
measurement was performed. It is clear that we can not extract
from equation (\ref{prob2}) any information about the final
observed value $b$ on region $B$. Therefore, the relational
approach is consistent with causality for a larger class of
operators, while the standard Bloch computation is extremely
restrictive as we have seen.\\ It is now clear that the relational
intrinsic order implies a new effect for partial causally
connected measurements in order to preserve the consistency with
causality. From a physical point of view, the intrinsic order
implies the suppression of the interference terms in equation
(\ref{prob1}) with respect to (\ref{probm2}). In other words,
quantum interference cannot arise among ``non-simultaneous"
alternatives. Notice however that (\ref{prob1}) coincides indeed
with (\ref{probm2}) for total or null causal connection.
\\ A final remark is in order, since as we said before we are not
taking into account the detailed description of the measurement
process. We have assumed the {\it minimally disturbing
hypothesis}. Under this hypothesis our approach is consistent with
causality in the general case without any need of introducing the
device into the system. It is frequently considered that, in order
to include non-local operators the {\it minimally disturbing
hypothesis} needs to be relaxed in order to describe a measurement
process avoiding causality violations within Bloch's
approach\cite{A-A,Preskill,Preskill2}. We have shown here that we
can conserve this hypothesis modifying instead Bloch's non
covariant order, and adopting therefore a covariant description.
The natural question to be asked is what are the physical
conclusions in connection to a possible extension of the
measurement process by including the devices degrees of freedom in
the relativistic case. If the minimally disturbing hypothesis is
experimentally consistent we should be able to understand this
suppression of interference terms once including the measurement
instrument within the theory in a fully covariant quantum
description. Besides decoherence associated to the whole non-local
$B$ measurement which gives us the diagonal density matrix in the
$b$-basis, we still need to understand why the coefficient are
calculated via (\ref{prob1}) and not (\ref{probm2}). It is clear
that this interference suppression related to the intrinsic
covariant order must rely on physical grounds. First of all notice
that the difference between (\ref{prob1}) and (\ref{probm2}) are
quantitively important in the case of partial causal connection
though both approaches coincide in the case of total or null
causal connection. It is therefore clear that there will be a
sudden change from the point of view of the probability calculus
from (\ref{probm2}) to (\ref{prob1}), as soon as the light cone
coming from $A$ connects region $B$, in order to keep a causal
behavior. It is possible to see that this change is continuous but
not smooth. It is naturally to ask to what sort of mechanism this
effect would be associated if one assumes now that the devices
degrees of freedom are included. A possible explanation could be
the following, let us suppose a smooth Schroedinger like wave
evolution for the combined system+apparatus, and a forward light
cone propagation, during the whole measurement process. Notice,
however, that the location of the region where measurement takes
place is still an external parameter in the theory, that is, there
is a fixed external causal structure. Then the discontinuities in
the derivative of the probability could arise due to the
appearance of quantum fields whose Schroedinger-like propagation
would affect one of the portions of $B$, and produce the above
effect as soon as the light cone coming from $A$ contacts region
$B$. In this case the microcausality property would start to play
an essential role and discontinuities would appear involving the
commutative properties of field operators. Since microcausality is
still an external property related to the fixed background it is
natural to associate this sort of ``phase transition" to the
relative change in the space-time position of the measurement
devices.\\

Summarizing, we have introduced a description of the relativistic
reduction process that gives new insight about the measurable
character of non-local operators in relativistic quantum
mechanics. Within our approach, the measurement of a wide class of
non-local operators does not break causality contrary to what
results from the non covariant approach. Further theoretical and
experimental efforts are required to understand these kind of
non-local measurements. It is clear by now that in order to
implement these type of measurements in practice, now thinking in
more realistic measurement instruments, it is fundamental to avoid
partial decoherence on $B_1,B_2$ which would suppress the
interference terms in (\ref{probm2}) leading to a causal behavior.
For the case we have studied above, we will need a quantum system
which specifically avoids a partial projection of $O_i$ due to
environmental interactions. Therefore, we need a macroscopic
physical system which behaves coherently during the measurement
process. The recent developments of SQUID (superconductor quantum
interference devices), where {\it macroscopic} states are put in
coherent superposition\cite{Fried,Legg}, may open the possibility
of an experimental verification. Let us sketch another possible
route: The operator introduced in (\ref{op}) could be associated
to the two point correlation function in a non translationally
invariant superconductor. In this system the two-point isothermal
susceptibility will be given by $\chi_T(x,y)=\beta G(x,y)$, where
$G(x,y)=<0|{\hat \phi}(x){\hat \phi}(y)|0>$ is the two-point Green
function for the effective quantum field associated with the order
parameter. Let us now calculate the net isothermal susceptibility
as $\chi_T=\frac{\partial M}{\partial H}$, with $M$ and $H$ the
magnetisation and magnetic field respectively. In the case of a
non translationally invariant system, i.e. $G(x,y)\neq G(x-y)$, we
will have $\chi_T=\beta\int_B\int_B dxdy G(x,y)$\cite{golden}.
Therefore, measuring global magnetic responses in superconductors
may open the possibility of introducing this sort of non-local
operators. An analogous reasoning could be made also for
superfluid systems. Similar measurements for fundamental fields
rather than effective quantum fields need further study.\\

Finally thinking about the possible physical applications of this
interference suppression mechanism is worthwhile to stress that it
occurs not directly related to environmental interaction but as a
consequence of a covariant order in the measurement process for
the case of partial causal connection\footnote{Environmental
decoherence is however expected to play an essential role during
the whole measurement process.}. These features may have important
consequences regarding quantum information processes. Hence, to
explore the experimental viability of these type of operators is
the crucial point to continue exploring along this path.\\

We wish to thank Hugo Fort, Jorge Pullin and Silvia Viola for many
helpful suggestions. We would like to thank also to the anonymous
referees for their insightful comments and suggestions which
certainly helped us to improve the ideas and presentation of the
manuscript.


\begin{thebibliography}{10}
\bibitem{A-A}Y.Aharonov and D.Albert, Phys. Rev. D 29, 228, (1984)
\bibitem{Sorkin} R.Sorkin, "Directions in General Relativity,
vol. II: a collection of Essays in honor of Dieter Brill's
Sixtieth Birthday" (CUP, 1993) B.L. Hu and T.A. Jacobson eds.
\bibitem{Preskill} D. Beckman, D.Gottesman, M.A. Nielsen and J.Preskill
Phys.Rev. A 64, 052309, (2001).

\bibitem{Preskill2} D. Beckman, D.Gottesman, A. Kitaev and J.Preskill Phys. Rev.D 65, 065022
(2002).
\bibitem{Bloch} I.Bloch, Phys. Rev. 156, 1377, (1967)
\bibitem{nos1} R. Gambini and R. A. Porto, Phys. Lett. A
Volume 294, Issue 3 (129-133), (2002).
\bibitem{zurek} For a recent review see W. Zurek, eprint: quant-ph/0306072 (2003)

\bibitem{nos2} R.Gambini and R.A. Porto, Phys. Rev. D 63, 105014, (2001).
\bibitem{nos3} R.Gambini and R.A. Porto,  New J. Phys. 4, 58,
(2002).
\bibitem{Esp}B.D'Espagnat. Physics Report 110, 202, (1984).
\bibitem{Tommas} Daniele Tommasini, JHEP 0207, 039, (2002).
\bibitem{Suarez} A. Stefanov, H. Zbinden, N. Gisin and A. Suarez, Phys. Rev. Lett. 88, 120404,
(2002) and reference therein.
\bibitem{Griff} Robert B. Griffiths, eprint quant-ph/0207015\\Robert B.
Griffiths, ``Consistent Quantum Theory" Cambridge University Press
(2002) ISBN: 0521803497.
\bibitem{Omnes} Roland Omnes, ``The Interpretation of Quantum
Mechanics", Princeton University Press (1994).
\bibitem{Fried} J.R. Friedman, V. Patel, W. Chen and J.E. Lukens,
Nature 406, 43 (2000).
\bibitem{Legg} J.A. Leggett, J. Phys. C 14, R415-451 (2002).
\bibitem{golden} Nigel Goldenfeld, ``Lectures on Phase Transitions
and the Renormalization Group", Perseus Books 1992.
\end{thebibliography}
\end{document}